\documentclass[conference]{IEEEtran}

\ifCLASSINFOpdf
\else
\fi
\usepackage{graphicx}
\usepackage{tabularx,booktabs}
\usepackage{dingbat}
\usepackage{diagbox}
\usepackage{multirow} 
\usepackage[hyphens]{url}
\usepackage[hidelinks,breaklinks]{hyperref}
\usepackage{xcolor}
\usepackage{amsfonts, amsmath, amsthm, amssymb}
\usepackage{subcaption}
\hypersetup{breaklinks=true}
\usepackage{tikz}
\usepackage{textcomp}
\usepackage{lipsum}
\IEEEoverridecommandlockouts
\newcommand\copyrighttext{%
  \footnotesize \textcopyright 2023 IEEE.  Personal use of this material is permitted.  Permission from IEEE must be obtained for all other uses, in any current or future media, including reprinting/republishing this material for advertising or promotional purposes, creating new collective works, for resale or redistribution to servers or lists, or reuse of any copyrighted component of this work in other works.}
\newcommand\copyrightnotice{%
\begin{tikzpicture}[remember picture,overlay]
\node[anchor=south,yshift=10pt] at (current page.south) {\fbox{\parbox{\dimexpr\textwidth-\fboxsep-\fboxrule\relax}{\copyrighttext}}};
\end{tikzpicture}%
}

\usepackage{graphicx}
\graphicspath{{figures/}}
\begin{document}

%
\title{Performance Evaluations of C-Band 5G NR FR1 (Sub-6 GHz) Uplink MIMO on Urban Train}

\author{
    \IEEEauthorblockN{Kasidis Arunruangsirilert\IEEEauthorrefmark{1}, Pasapong Wongprasert\IEEEauthorrefmark{2}, Jiro Katto\IEEEauthorrefmark{1}}
    \IEEEauthorblockA{\IEEEauthorrefmark{1}Department of Computer Science and Communications Engineering, Waseda University, Tokyo, Japan}
    \IEEEauthorblockA{\IEEEauthorrefmark{2}International School of Engineering, Chulalongkorn University, Bangkok, Thailand
    \\\{kasidis, katto\}@katto.comm.waseda.ac.jp, Pasapong.Wongprasert@gmail.com}
}
%
\maketitle
\copyrightnotice
\begin{abstract}
Due to the recent demand for huge Uplink throughput on Mobile networks driven by the rapid development of social media platforms, UHD 4K/8K video, and VR/AR contents, Uplink MIMO (UL-MIMO) has now been deployed on commercial 5G networks with reasonable availability of supported User Equipment (UE) for consumers. By utilizing up to 2 Tx antenna ports, UL-MIMO-capable UE promised to achieve up to two times the uplink throughput in ideal conditions, while providing improved uplink performance over UE with 1Tx in challenging conditions.

In Japan, SoftBank, one of the carriers, introduced 5G Standalone (SA) services for the Fixed Wireless Access (FWA) application back in October 2021. Mobile services were commenced in May 2022, which provide UL-MIMO for supported UE on C-Band or Band n77 (3.7 GHz). In this paper, the uplink performance of UL-MIMO-capable UE will be compared against the conventional UL-1Tx UE on trains, which is the most popular method of transportation for the Japanese. The results show that UL-MIMO-capable UE delivers an average of 19.8\% better throughput on moving trains with up to 33.5\% in the more favorable signal conditions. A moderate relationship between downlink 5G NR SS-RSRP and uplink throughput also has been observed.
\end{abstract}



%
\IEEEpeerreviewmaketitle

\section{Introduction}
While Uplink MIMO (UL-MIMO) was introduced back during the 4G era, where it has been defined in the 3GPP TS 36.817 Release 10 from 2010 \cite{3GPP_36-817}, there has been little to no implementation in the commercial environment. With the introduction of 4G Long-Term Evolution (LTE) and Downlink MIMO (DL-MIMO) in the 2010s, Network operators around the world focused on improving the downlink performance of their network, which is heavily limited in 3G Universal Mobile Telecommunications System (UMTS) technology. Later, the introduction of Time-Division Duplexing LTE (TDD-LTE), which allows network operators to have flexible control over the amount of downlink and uplink slots over time, further emphasized the importance of downlink traffic. Major mobile network operators like NTT DoCoMo configured their TDD network with a 7:2 downlink to uplink ratio, allocating the majority of resources toward downlink \cite{sagae_sawamukai_ohwatari_kiyoshima_kanbara_takahashi_2020} to cope with the rapidly increasing demand for consumption of multimedia content over the mobile network. Uplink performance was not a concern at the time because smartphone technology was not mature enough for large-scale multimedia content creation on the go. The high-end smartphone system-on-chips (SoC) at the time such as Qualcomm Snapdragon S4, released in early 2012, was only capable of 1080p video encoding \cite{qualcomm_2012}. Therefore, most uplink traffic on mobile networks was the transmission of email, instant messages, photos, and standard-quality video, which does not require large uplink throughput to accomplish. At the time, most large-scale multimedia content production would use a fixed broadband connection, and news gathering from a remote location would utilize satellite links to relay the video feed back to TV stations.

In the late 2010s, the performance of smartphone SoCs had improved significantly with the high-end SoCs like Qualcomm Snapdragon 855 now capable of encoding 4K videos in real-time with the more efficient codecs such as High Efficiency Video Coding (HEVC) \cite{qualcomm_855}. Furthermore, social network platforms like Facebook \cite{constine_2015} and Instagram \cite{constine_2016} also launched live streaming services at around the same time, which opens up many possibilities for content creation on the go. Advancement in broadcast technology has brought portable TV encoders to the market, which utilized mobile networks for the uplink instead of conventional satellite links with BBC news first transitioning to LiveU's cellular uplink technology as early as 2010 \cite{published_2012}. As more and more content creation tasks now rely on the uplink of mobile networks, the demand for uplink on mobile networks began to take off.

With the introduction of 5G, came the new use cases that will put more strain on the uplink of mobile networks including the live streaming of Ultra High Definition (UHD) 4K/8K Virtual Reality (VR)/Augmented Reality (AR) content, the video feed from multiple cameras on Self-Driving Vehicles, Internet of Things, Industrial Automation, etc \cite{9868877}. Due to the increasing demand for high uplink performance, UL-MIMO has been made available commercially for the first time on a 5G network with healthy support from both network operators and device manufacturers \cite{goovaerts_2021}. While almost all devices that support the 5G Frequency Range 2 (FR2 or mmWave) band also have support for UL-MIMO on the 5G FR2 (mmWave) band, not all devices that support the 5G Frequency Range 1 (FR1 or Sub-6 GHz) band actually support UL-MIMO on all of their supported 5G FR1 (Sub-6 GHz) frequency bands, with some devices lacking UL-MIMO capability completely. This can result in massive differences in uplink performance between different models of UE. As more and more users continued to upload larger and larger content to the internet via the mobile network on the go, the uplink performance on mobile networks will gradually become the next point of interest. In this paper, the uplink performance of UL-MIMO-capable UE will be evaluated and compared against the conventional UL-1Tx UE. The uplink performance while moving on the train in the Tokyo Metropolitan area will be the main point of discussion as trains are the most popular method of transportation in Tokyo, Japan. This paper is organized as follows. Section II will discuss the experiment environment. Section III will provide experimental results as well as an analysis of the result. Finally, the conclusion and future work shall be discussed in Section IV.

\section{Experiment Environment}

\subsection{User Equipment (UE)}
\begin{figure}[t!]
  \centering
  \includegraphics[width=0.37\textwidth]{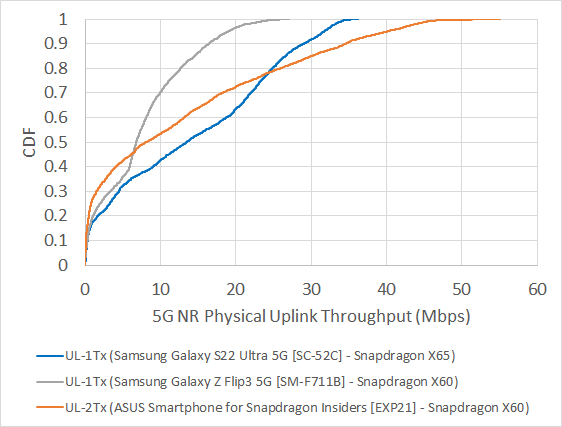}
  \setlength{\belowcaptionskip}{-18pt}
  \caption{CDF of Physical Uplink Throughput on Various UE}
  \label{fig:UECDF}
  
\end{figure}

For the User Equipment (UE), ASUS Smartphone for Snapdragon Insiders (EXP21) with Qualcomm Snapdragon X60 5G RF Modem, which has UL-2Tx and supports UL-MIMO on 5G band n77, was used. Similar to the base station in downlink MIMO \cite{ericsson} configuration, UE with UL-MIMO capability may operate in several Tx modes; single-layer transmission using two antenna ports (Transmit Diversity), two-layer transmission using two antenna ports with transform precoding disabled (Spatial Multiplexing), and single-layer transmission on a single antenna port (UL-1Tx); depending on the channel condition \cite{3GPP_38-211}. UL-MIMO-capable UE, which has two antenna ports, may also be generally referred to as UL-2Tx, regardless of the operating Tx mode.

The use of different UE to obtain UL-1Tx and UL-2Tx results was also considered. However, modem chipset and antenna design vary from one smartphone to another, which may add more variables to the experiment results. This was confirmed by running three trials of the experiment on three different UE, as seen in Figure \ref{fig:UECDF}. Therefore, the decision was made to only use ASUS Smartphone for Snapdragon Insiders (EXP21), then modify the modem firmware to disable one of the two antennae to obtain UL-1Tx result. By doing this, the variable caused by differences in antenna designs, RF amplification circuits, and modem chipset can be eliminated.

By using NSG, the UECapabilityInformation packet was captured and the capability of two configurations were verified. The uplink throughput performance of both configurations was also measured in front of the base station with the channel bandwidth of 100 MHz to confirm its capability. In UL-MIMO configuration, peak physical uplink throughput of 206.99 Mbps was measured, compared to 109.85 Mbps of UL-1Tx counterpart, demonstrating that UL-MIMO can deliver almost two times the uplink performance in the ideal condition. Additionally, maximum theoretical uplink throughput was also calculated based on 3GPP TS 38.306 standard \cite{3GPP_38-306}. Parameter for calculation based on the actual configuration of SoftBank as well as the result can be seen in Table \ref{tab:TheoryThpt}.

\begin{table}[!tbp]
\caption{5G Uplink Theoretical Throughput Calculation Parameters and Results}
\centering
\label{tab:TheoryThpt}
\resizebox{6cm}{!}{\begin{tabular}{@{}lll@{}}
\toprule
Parameter                       & UL-2Tx   & UL-1Tx   \\\midrule
Frequency Band                  & n77       & n77       \\
Channel Bandwidth (MHz)         & 100       & 100       \\
Sub Carrier Spacing (SCS)       & 30 kHz    & 30 kHz    \\
TDD Pattern 1 Periodictity (ms) & 3 & 3\\
TDD Pattern 1 Slots (DL/UL)     & 3/2 & 3/2\\
TDD Pattern 1 Symbols (DL/UL)   & 6/4 & 6/4\\
TDD Pattern 2 Periodictity (ms) & 2 & 2\\
TDD Pattern 2 Slots (DL/UL)     & 4/0 & 4/0\\
TDD Pattern 2 Symbols (DL/UL)   & 0/0 & 0/0\\
Uplink Modulation               & 256QAM & 256QAM\\
Uplink Layer                    & 2 & 1\\
Scaling Factor                  & 1.0 & 1.0\\
Maximum Throughput (Mbps)       & 285.72 & 142.86\\
\bottomrule
\end{tabular}}
\vspace{-6mm}
\end{table}

Since the test results show that the peak uplink throughput of ASUS Smartphone for Snapdragon Insiders with UL-2Tx and UL-MIMO enabled exceeded the maximum theoretical throughput of single transmit UE by about 45\%, UL-MIMO was verified to be compatible and usable in the experimental environment.

\subsection{Modulation Scheme Calculation}

Typically, UE will report the maximum modulation scheme that it's currently using. However, in 5G, each resource block can be modulated using a different modulation scheme, so using the maximum modulation scheme would not provide a good representation of the channel condition. The software used can provide the utilization percentage of each modulation scheme, so the formula used to find the average modulation order for each sample can be defined as follows:

\begin{equation} \label{eq:1}
Average Modulation Order = \sum_{m=1}^{M} P_m Q_m
\end{equation}

Given \begin{math}M\end{math} modulation schemes available, \begin{math}P_m\end{math} represents the utilization percentage of \begin{math}Modulation_m\end{math}, and \begin{math}Q_m\end{math} represents the modulation order of such modulation scheme.

\subsection{Network Environment} \label{netenv}

As the majority of consumer UE only support 5G FR1 (Sub-6 GHz) UL-MIMO in 5G Standalone (SA) mode, a 5G SA network is required for the experiment. At the time of research, only SoftBank provided 5G SA service in Tokyo, Japan. Therefore, SoftBank was chosen for the evaluation in this paper. Measurements show that SoftBank deploys two carriers of 5G in frequency band n77. The first carrier is located at 3.40-3.44 GHz with a channel bandwidth of 40 MHz, and the second carrier is located at 3.90-4.00 GHz with a channel bandwidth of 100 MHz. It has been found that the second carrier, with higher carrier frequency, is only being deployed in high-density crowded environments, likely due to higher path loss from higher frequencies making them less desirable for providing wide coverage.

Throughout the experiment, 90\% of the data was collected on the 40 MHz wide carrier and the remaining 10\% was collected on the 100 MHz wide carrier. Most of the evaluation will be performed on the data obtained from the 40 MHz wide carrier as the sample size is significantly larger than the 100 MHz counterpart. However, some of the data from the 100 MHz carrier will also be included for reference purposes.

\begin{figure}[t!]
\centering
\begin{subfigure}{.24\textwidth}
  \centering
  \includegraphics[width=0.95\linewidth]{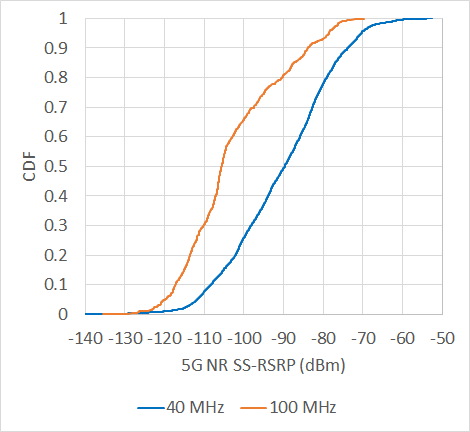}
  \caption{CDF of SS-RSRP}
  \label{fig:sub51}
\end{subfigure}%
\begin{subfigure}{.24\textwidth}
  \centering
  \includegraphics[width=0.95\linewidth]{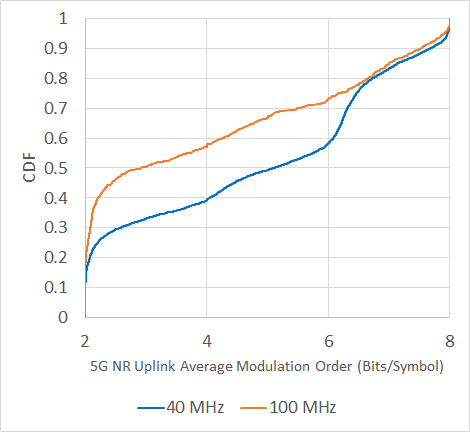}
  \caption{CDF of Average Mod. Order}
  \label{fig:sub52}
\end{subfigure}\\
\begin{subfigure}{.24\textwidth}
  \centering
  \includegraphics[width=0.95\linewidth]{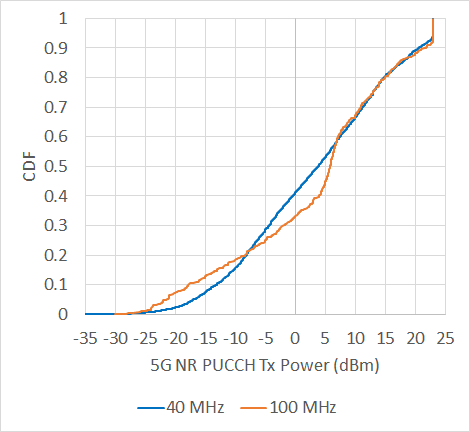}
  \caption{CDF of PUCCH Tx Power}
  \label{fig:sub53}
\end{subfigure}%
\begin{subfigure}{.24\textwidth}
  \centering
  \includegraphics[width=0.95\linewidth]{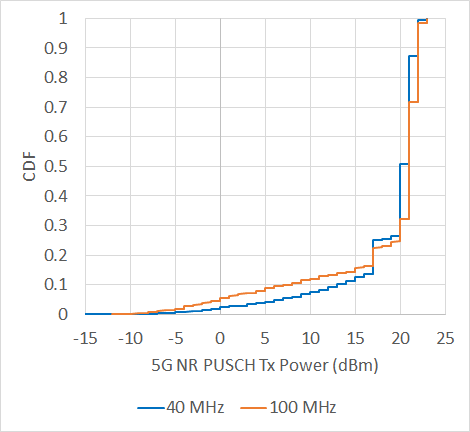}
  \caption{CDF of PUSCH Tx Power}
  \label{fig:sub54}
\end{subfigure}
\setlength{\belowcaptionskip}{-10pt}
\caption{Summary of RF Parameters for each carrier}

\end{figure}
\begin{table}[!tbp]
\caption{5G Uplink Theoretical Throughput Calculation for Various Channel Bandwidths}
\centering
\label{tab:40MHzCalc}
\resizebox{7cm}{!}{\begin{tabular}{@{}lll@{}}
\toprule
Parameter                       & UL-2Tx   & UL-1Tx   \\\midrule
Maximum Throughput - 40 MHz (Mbps)       & 110.94 & 55.47\\
Maximum Throughput - 100 MHz (Mbps)       & 285.72 & 142.86\\
\bottomrule
\end{tabular}}
\vspace{-6mm}
\end{table}

As seen in Figure \ref{fig:sub51}, the measured SS-RSRP from the 100 MHz carrier is significantly weaker than the 40 MHz carrier, which is to be expected as the 100 MHz carrier is being deployed as small cells to increase the capacity. Since the p-max parameter was configured by the network operator to be at 23 dBm, the maximum power that the UE can transmit was limited to 23 dBm. The transmission power of the Physical Uplink Control Channel (PUCCH) represents the power used to transmit signaling and control information from UE to the base station, which does not affect by load or user activity. Typically, UE will increase its PUCCH Tx Power as needed to reach the base station, and communication may be difficult when PUCCH Tx Power is at maximum value as it cannot increase them any further. The figure \ref{fig:sub53} shows that the distribution of PUCCH Tx Power used for both carriers displayed similar characteristics, which is likely due to the weaker average SS-RSRP of the 100 MHz cell causing UE to handover away from the 100 MHz cell before the PUCCH Tx Power became an issue.

On the other hand, the Tx Power of Physical Uplink Shared Channel (PUSCH) may vary depending on the user activity and the amount of user data that needed to be transmitted from the UE to the base station. Because the stress test was performed throughout the experiment, PUSCH Tx power is usually at the maximum unless the condition is exceptionally ideal as the UE tried to use the highest modulation scheme possible to maximize the uplink throughput and only step down the power when the maximum modulation scheme is reached. Similar to PUCCH Tx Power, Figure \ref{fig:sub54} shows a similar distribution between two carriers with UE spending more than 70\% of the time using PUSCH Tx Power more than 20 dBm. However, the modulation index shown in Figure \ref{fig:sub52} shows that the average modulation scheme used for PUSCH for the 100 MHz carrier is significantly worse than the 40 MHz carrier, which shows that the uplink channel condition of the 100 MHz is significantly worse than 40 MHz. This is likely caused by the lower cell density of the 100 MHz deployment and lower maximum Tx power per resource block for the 100 MHz as a 100 MHz channel with SCS of 30 kHz contains 273 resources blocks, whereas a 40 MHz channel with the same configurations contains only 106 resource blocks.

For evaluation accuracy, the uplink performance on each carrier will be filtered and evaluated separately. Additionally, the maximum theoretical throughput for 40 MHz was also calculated and shown in Table \ref{tab:40MHzCalc} as reference. It should be noted that the rest of the parameters used for the calculation aside from channel bandwidth remains the same as Table \ref{tab:TheoryThpt}.

\subsection{Test Route and Data Collection}
\begin{figure}[t!]
  \centering
  \includegraphics[width=0.40\textwidth]{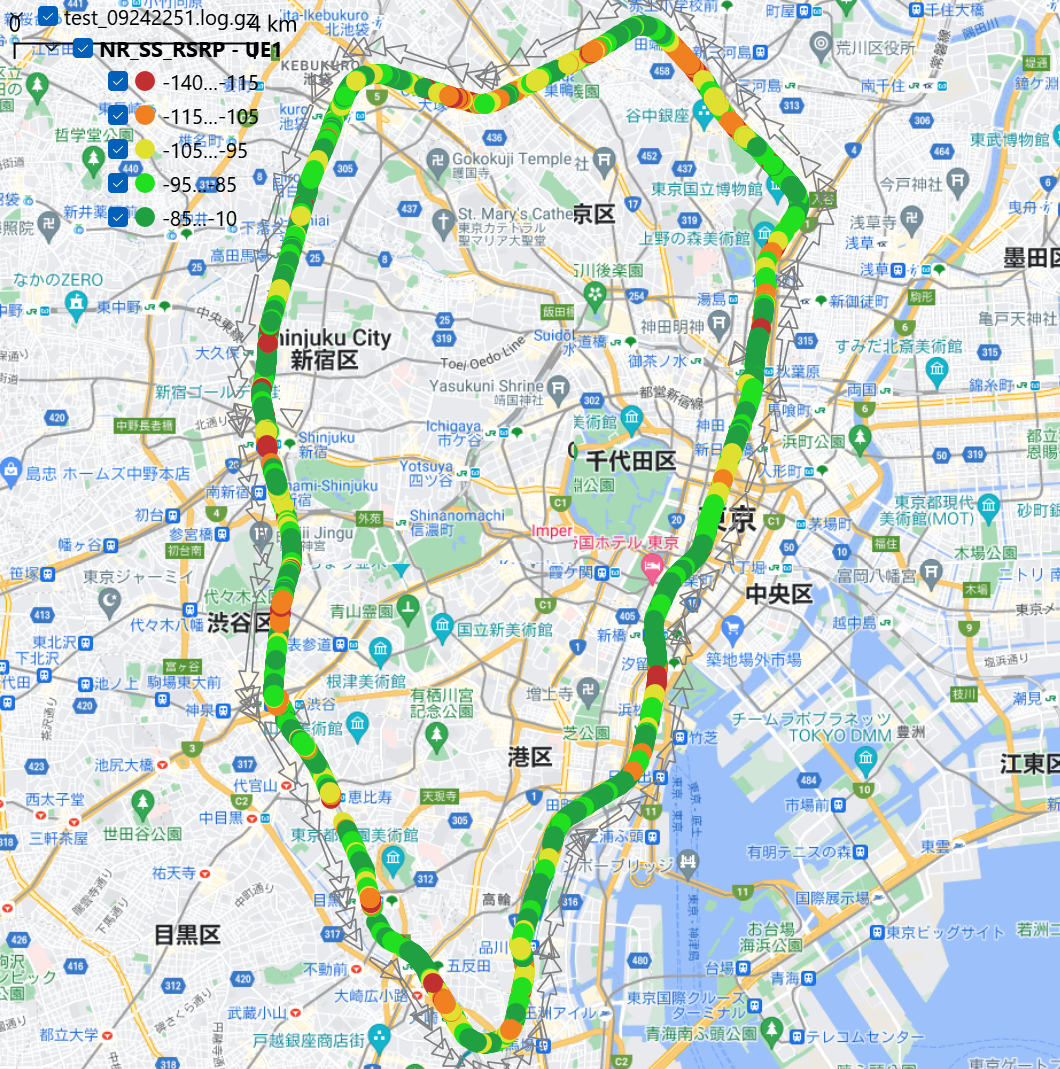}
  \setlength{\belowcaptionskip}{-16pt}
  \caption{Map of Test Route on JR Yamanote Line, the color represented 5G SS-RSRP}
  \label{fig:RSRP-Map}
\end{figure}
\begin{figure}[t!]
  \centering
  \includegraphics[width=0.36\textwidth]{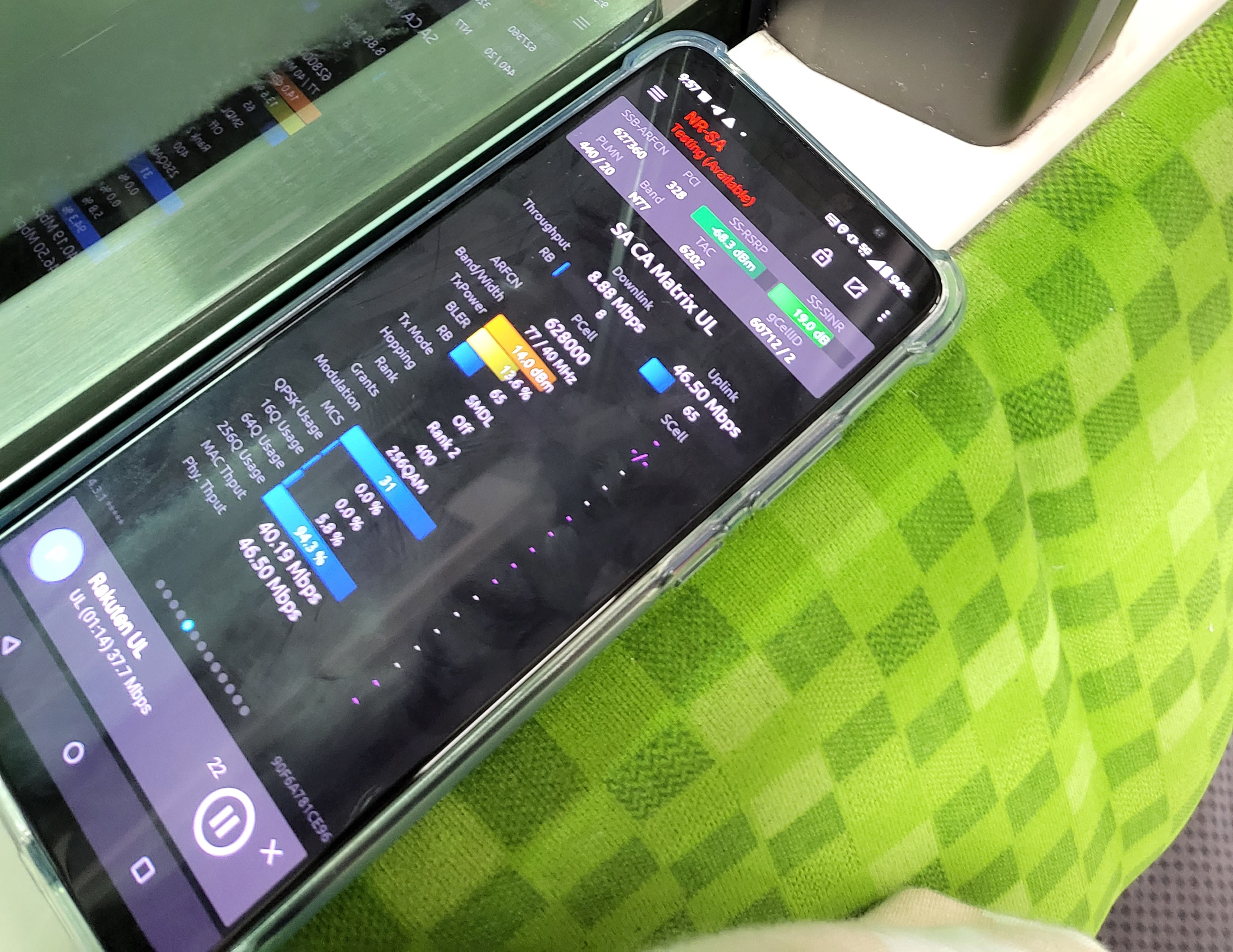}
  \setlength{\belowcaptionskip}{-13pt}
  \caption{Test Setup}
  \label{fig:TestEnv}
\end{figure}
\begin{figure}[t!]
  \centering
  \includegraphics[width=0.42\textwidth]{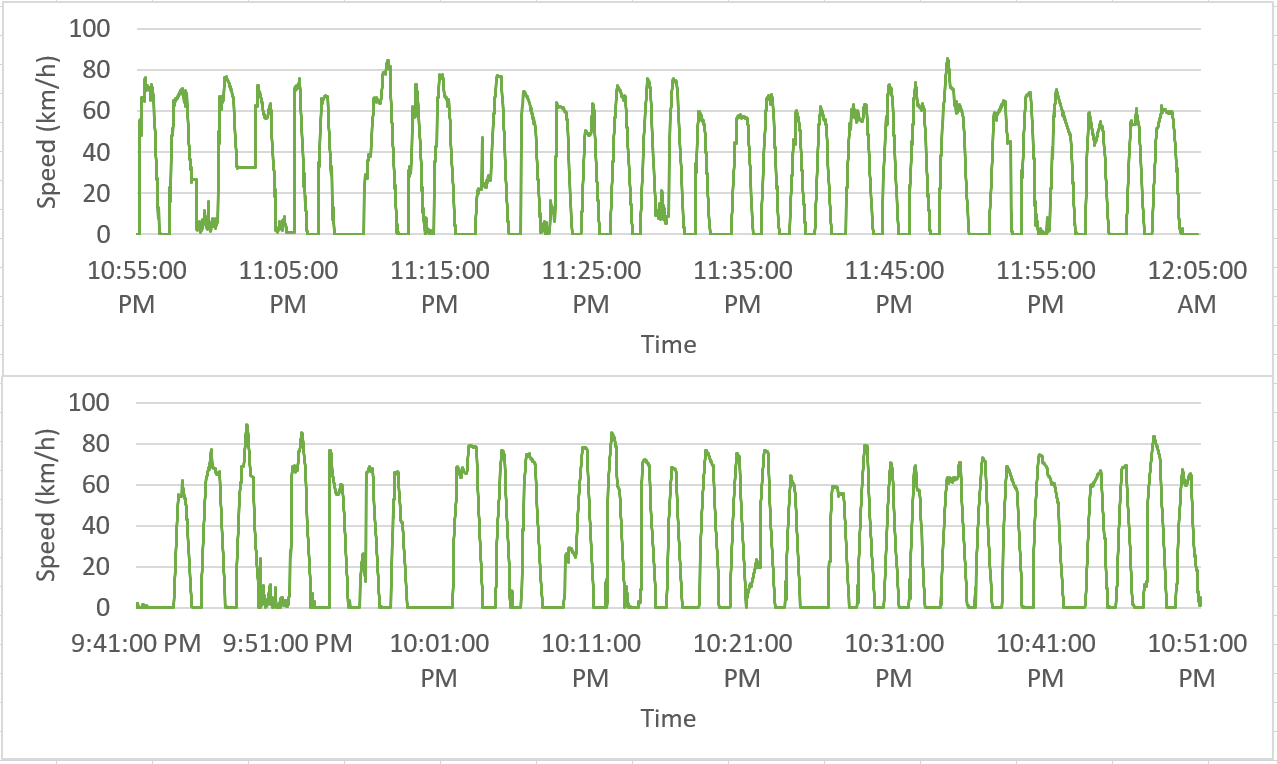}
  \setlength{\belowcaptionskip}{-21pt}
  \caption{Mobility Characteristics on JR Yamanote Line}
  \label{fig:Mobility}
\end{figure}
One of the most used urban train lines in Tokyo is the JR Yamanote Line, the loop line in the middle of the city serving Tokyo's major stations and city centers. The line is 34.5 km long with 30 stations along the way. Since the whole line is located in central Tokyo, there were many 5G base stations deployed densely throughout the route, which should be a good representation of typical 5G deployment in an urban area. The map of the route can be seen in Figure \ref{fig:RSRP-Map}.

Data collection was done by placing the UE on the side window of the train as seen in Figure \ref{fig:TestEnv} to ensure the best line of sight to the base station (gNodeB). The experiment was conducted between 21:30 to 00:30 to ensure minimal base station load, which can affect the experiment. One trial, which involved looping the Yamanote Line once, was performed for each configuration. The data was collected using a cellular network drive test tool called "Network Signal Guru (NSG)," which allows the in-depth result collection of RF parameters, throughput, and signaling log. For the uplink throughput experiments, the test function of NSG was used to perform a continuous upload stress test via HTTP POST to the test server. The timeout detection was set to 10 seconds, which means that if the UE failed to transmit the data for 10 seconds, then it will attempt to re-establish the HTTP connection. On the other hand, the latency and packet loss experiments were also performed by using the test function of NSG. Pings were sent to the target server, Google DNS (8.8.8.8), then response time was measured for every request. If the response time exceeds 4000ms, the packet is considered to be lost, and if the response time exceeds 100ms, then the packet is considered to be a high latency packet.

While the best RF parameter to use for the evaluation is the RSRP at the base station side, which is the actual power that the base station received from the UE, the data is considered confidential and only accessible by the network operator. Another alternative is Timing Advance (TA), which is used by the UE to adjust its uplink transmission so that the downlink and uplink subframes are synchronized at the base station side. However, TA only updates during the RACH procedure, which only happens during attach, handover, etc. Therefore, PUCCH Tx Power and SS-RSRP will be used for the evaluation. PUCCH Tx Power can help determine the difficulty of UE uplink reaching the base station without accounting for the user load as explained in Section \ref{netenv}, while SS-RSRP is typically used for cell reselection and handover.

\section{Results and Analysis}

\subsection{Physical Uplink Throughput}

The first trial started at Shinjuku Station at 21:41 with UL-MIMO enabled on the UE. One counter-clockwise was taken on JR Yamanote Line and the experiment was finished at Shinjuku Station at 22:51. Since the experiment needed to be done at off-peak hours to prevent network congestion, the UE was reprogrammed on the train to disable the UL-MIMO, leaving only one Tx enabled, then the second trial was started at 22:55 at Harajuku Station, one full counter-clockwise loop was taken, then the experiment was concluded at Harajuku Station at 00:05. The mobility characteristics as measured by the GPS can be seen in Figure \ref{fig:Mobility}.
\begin{table}[!tbp]
\caption{Summary of 5G NR Physical Uplink Throughput by Configuration}
\centering
\label{tab:Results}
\resizebox{7.5cm}{!}{\begin{tabular}{@{}llllllll@{}}
\toprule
Configuration      & BW & 5\% & 25\% & Median & 75\% & 95\% & Average  \\ \midrule
UL-1Tx & 40&0.07&1.89&9.22&20.97&29.40&11.54\\
UL-2Tx &&0.04&1.80&9.95&23.08&39.25&13.83\\
\midrule
UL-1Tx &100&0.11&0.80&6.87&12.96&35.88&9.86\\
UL-2Tx &&0.06&0.86&3.33&14.30&37.00&9.40\\
\bottomrule
\end{tabular}}
\vspace{-3mm}
\end{table}
\begin{figure}[t!]
\centering
\begin{subfigure}{.24\textwidth}
  \centering
  \includegraphics[width=0.95\linewidth]{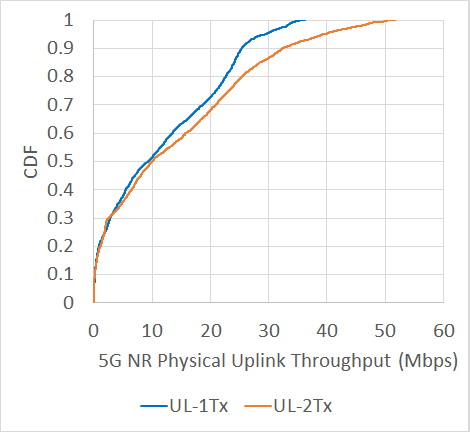}
  \caption{40 MHz Wide Carrier}
  \label{fig:sub81}
\end{subfigure}%
\begin{subfigure}{.24\textwidth}
  \centering
  \includegraphics[width=0.95\linewidth]{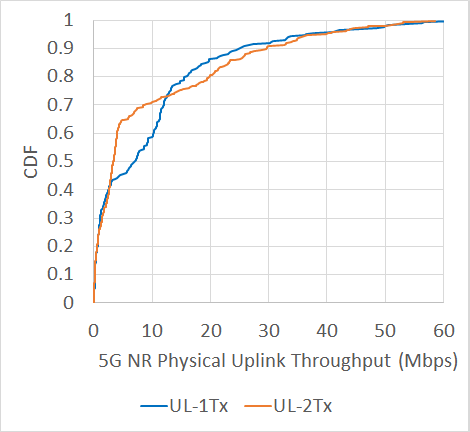}
  \caption{100 MHz Wide Carrier}
  \label{fig:sub82}
\end{subfigure}
\setlength{\belowcaptionskip}{-20pt}
\caption{CDF of 5G NR Uplink Physical Throughput}
\end{figure}

The difference between UL-1Tx and UL-2Tx is more pronounced as the signal condition became more favorable, as the experiment results show that while the experiment results show in Table \ref{tab:Results} shows that the UL-MIMO provides a 19.8\% improvement in the physical uplink throughput compared to the UL-1Tx on the 40 MHz carrier, the difference when the UE is on the 100 MHz carrier is negligible. On the 40 MHz carrier, the difference in median value was negligible, however, the difference grows bigger at the third quartile and 95 percentile, with UL-MIMO, delivering the improvement of 10\% and 33.5\%, respectively. While the throughput performance on the 40 MHz carrier of UL-MIMO began to divert from UL-1Tx above the median as shown in Figure \ref{fig:sub81}, the poor coverage of the 100 MHz carrier caused the signal quality to be poor most of the time, so UL-MIMO provides a negligible improvement on the 100 MHz carrier. Therefore, results show that the throughput improvement from UL-MIMO can only be achieved with adequate cell density and reasonably good coverage.

\begin{figure}[t!]
\centering
\begin{subfigure}{.24\textwidth}
  \centering
  \includegraphics[width=0.95\linewidth]{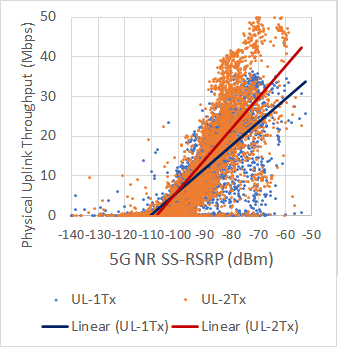}
  \caption{By SS-RSRP}
  \label{fig:sub71}
\end{subfigure}%
\begin{subfigure}{.24\textwidth}
  \centering
  \includegraphics[width=0.95\linewidth]{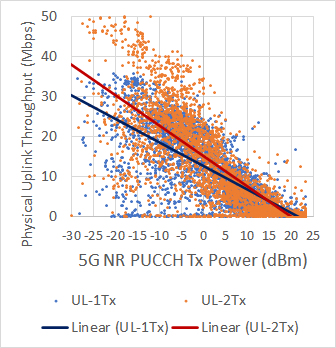}
  \caption{By PUCCH Tx Power}
  \label{fig:sub72}
\end{subfigure}
\setlength{\belowcaptionskip}{-15pt}
\caption{Physical Uplink Throughput on 40 MHz carrier}
\end{figure}
\begin{figure}[t!]
\centering
\begin{subfigure}{.20\textwidth}
  \centering
  \includegraphics[width=0.7\linewidth]{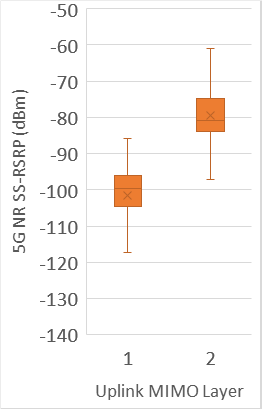}
  \caption{By SS-RSRP}
  \label{fig:sub101}
\end{subfigure}%
\begin{subfigure}{.20\textwidth}
  \centering
  \includegraphics[width=0.7\linewidth]{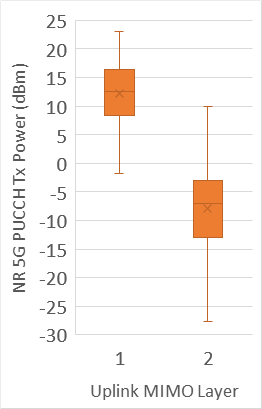}
  \caption{By PUCCH Tx Power}
  \label{fig:sub102}
\end{subfigure}
\setlength{\belowcaptionskip}{-20pt}
\caption{Uplink MIMO Layer utilized on 40 MHz carrier}
\end{figure}

The experiment results also demonstrated a moderate relationship for both between the physical uplink throughput and the downlink SS-RSRP and between the physical uplink throughput and 5G PUCCH Tx Power as shown in Figure \ref{fig:sub71} and \ref{fig:sub72}. When considering the SS-RSRP from Figure \ref{fig:sub71}, it has been found that UL-MIMO provides no improvement when the SS-RSRP is below -101.9 dBm. While the improvement of 10\% and 20\% can be observed at the relatively weak signal of -98.4 dBm and -89.9 dBm, respectively. At the SS-RSRP of -75 dBm, which can be easily obtainable with a direct line of sight to the base station, an improvement of 26\% was observed. The relationship between the uplink throughput and PUCCH Tx Power also displayed gradually larger improvements as the UE is using lower Tx power for the uplink control channel, which demonstrated that as the uplink channel quality became better, UL-MIMO effectiveness improved.

\begin{figure}[t!]
\centering
\begin{subfigure}{.24\textwidth}
  \centering
  \includegraphics[width=0.96\linewidth]{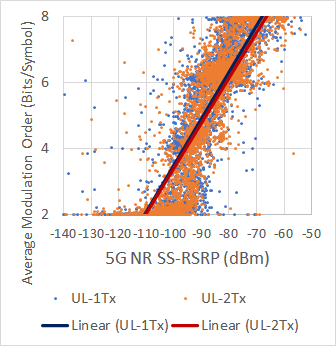}
  \caption{By SS-RSRP}
  \label{fig:ULMod}
\end{subfigure}%
\begin{subfigure}{.24\textwidth}
  \centering
  \includegraphics[width=0.96\linewidth]{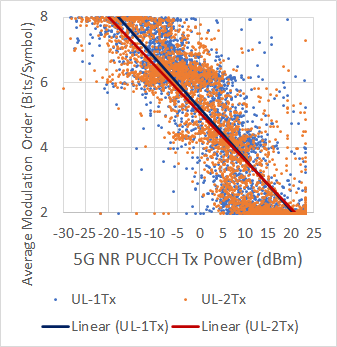}
  \caption{By PUCCH Tx Power}
  \label{fig:sub112}
\end{subfigure}
\setlength{\belowcaptionskip}{-5pt}
\caption{Average Uplink Modulation Order on 40 MHz carrier}
\end{figure}
\begin{figure}[t!]
\centering
\begin{subfigure}{.24\textwidth}
  \centering
  \includegraphics[width=0.95\linewidth]{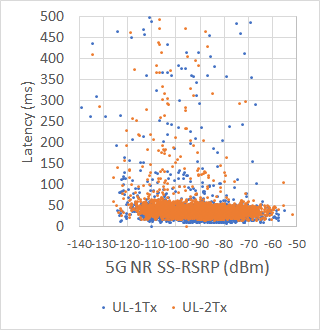}
  \caption{Latency vs SS-RSRP}
  \label{fig:sub121}
\end{subfigure}%
\begin{subfigure}{.24\textwidth}
  \centering
  \includegraphics[width=0.95\linewidth]{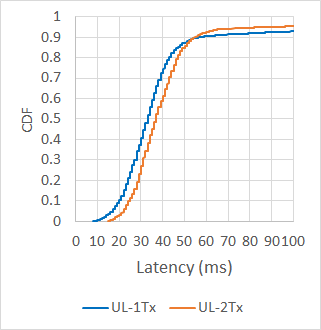}
  \caption{CDF of Latency}
  \label{fig:sub122}
\end{subfigure}
\setlength{\belowcaptionskip}{-18pt}
\caption{Summary of Latency Performance}
\end{figure}

While UL-MIMO-capable UE has an ability to use two antennae to transmit different two independent data streams to improve the throughput in spatial multiplexing transmission mode, it also has the ability to transmit the same data stream to improve coverage at a cell edge in transmit diversity transmission mode. Utilization of each transmission mode by RSRP can be seen in Figure \ref{fig:sub101}, which shows that the UE choose to utilize transmit diversity mode at an average SS-RSRP of -101.7 dBm or higher and spatial multiplexing mode at an average SS-RSRP of -79.5 dBm or higher while switching between two modes as channel condition improved or degraded in the intermediate SS-RSRP between the two average values. This explains the gradual improvement in throughput starting from SS-RSRP of -101.9 dBm as discussed earlier. When considering the PUCCH Tx Power, Figure \ref{fig:sub102} also shows similar results with spatial multiplexing being used when the channel condition is more favorable.

The transmit diversity transmission mode promised to improve uplink performance by reducing interference by using two antennae for one layer uplink transmission, allowing for higher modulation order at weaker signal conditions. However, the experiment results shown in Figure \ref{fig:ULMod} demonstrate the negligible difference between UL-1Tx and UL-MIMO-capable UE, which means most of the throughput gain was the direct result of the operation in spatial multiplexing mode.

\subsection{Latency and Packet Loss}

\begin{figure}[t!]
\centering
\begin{subfigure}{.2412\textwidth}
  \centering
  \includegraphics[width=1\linewidth]{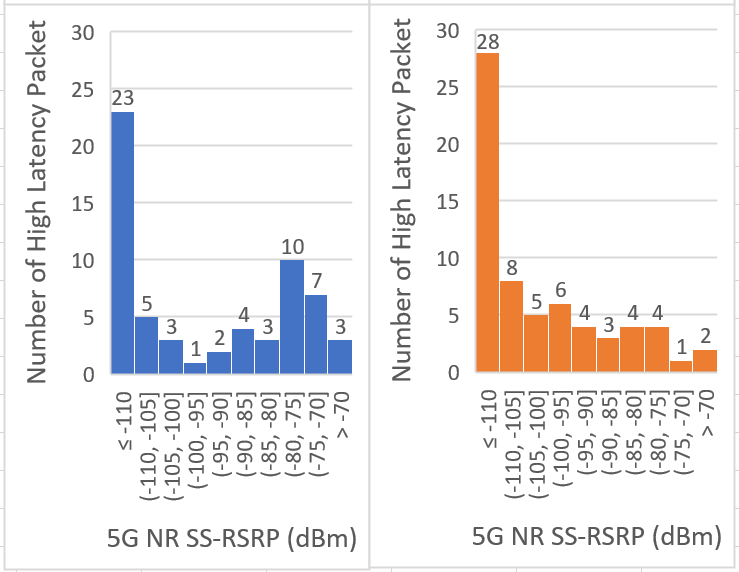}
  \caption{High Latency Packet}
  \label{fig:sub131}
\end{subfigure}%
\begin{subfigure}{.2412\textwidth}
  \centering
  \includegraphics[width=1\linewidth]{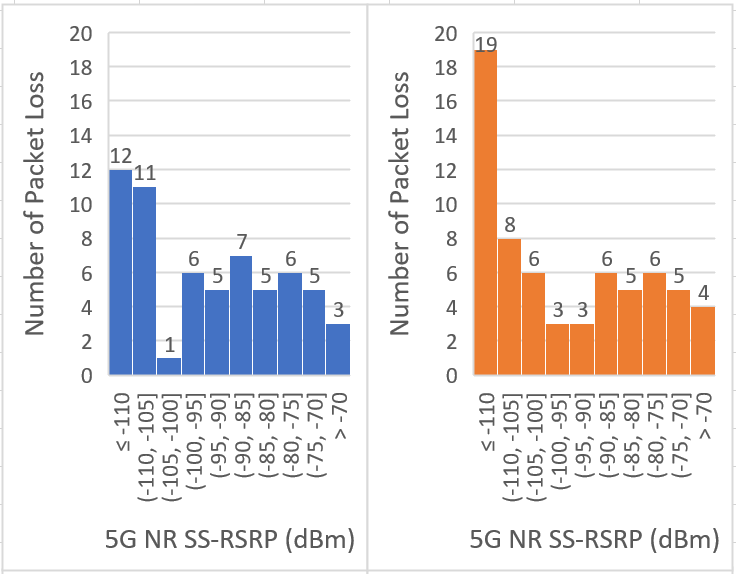}
  \caption{Packet Loss}
  \label{fig:sub132}
\end{subfigure}
\setlength{\belowcaptionskip}{-18pt}
\caption{Summary of Packet Loss and High Latency Packet (Blue = UL-1Tx, Orange = UL-2Tx)}
\end{figure}

From the experiment results, it has been found that the latency performance with UL-MIMO enabled is slightly worse than the UL-1Tx counterpart. UL-1Tx delivers 4ms lower latency compared to UL-MIMO regardless of the signal condition. While the distribution of packet loss and high latency packet is similar between UL-MIMO and UL-1Tx configurations as seen in Figure \ref{fig:sub131} and \ref{fig:sub132}, in poor signal conditions, the number of packet loss and high latency packets of UL-1Tx configuration is significantly lower than UL-MIMO counterpart likely due to the difference in the maximum PUSCH power allowed per antenna. As the p-max value is 23 dBm, UL-1Tx UE is allowed to use the full 23 dBm on one antenna, whereas UL-MIMO UE is only allowed to transmit 20 dBm per antenna, which may allow UL-1Tx UE to perform slightly better in very weak signal condition. This cause the UL-1Tx UE to drop 37\% less packet and yield 18\% less high latency packet when the SS-RSRP is weaker than -110 dBm.

\section{Conclusions and Future Work}
In this paper, the uplink performance of UL-MIMO-capable 5G User Equipment (UE) was evaluated against the conventional UL-1Tx implementation on a live commercial network while moving on urban trains in the Tokyo Metropolitan area. The result shows that UL-MIMO-capable UE provides an average improvement of 19.8\% improvement in physical uplink throughput compared to the single transmit counterpart. Larger improvement was observed as the signal condition became more favorable as UL-MIMO UE can spend more time on spatial multiplexing transmission mode, which allows up to two layers of the data stream to be transmitted with up to 33.5\% improvement observed when comparing 95 percentile results between each uplink configuration. The results also show that the UE with UL-MIMO is likely to utilize the spatial multiplexing transmission mode when SS-RSRP is above -80 dBm and PUCCH Tx power of lower than -5 dBm.

While the difference in latency and packet loss performance between the two configurations in typical use cases is minor, it has been found that UL-1Tx did delivers 4ms lower latency in most signal conditions. However, in very poor signal conditions, UL-1Tx performed significantly more stable than UL-MIMO yielding 37\% less packet loss and 18\% less high latency packet when SS-RSRP is lower than -110 dBm.

Therefore, network operators should consider increasing the base station density during network deployment, so that the UE with UL-MIMO can enjoy a significantly improved experience, and enable new use cases of 5G by providing good quality of experience to uplink heavy tasks, such as live streaming of UHD 4K/8K video, transmission of live multi-camera video feed from self-driving vehicles, and uplink of telemetry data in industrial automation scenarios, while reducing the possibility of poor latency performance of UL-MIMO-capable UE in poor signal condition.

As for future work, the moderate relationship between the downlink SS-RSRP, PUCCH Tx Power, and physical uplink throughput will be utilized for the network performance prediction to improve users' quality of experience when streaming UHD 4K/8K video in challenging RF conditions to prevent re-buffering, while preventing under-utilization of available bandwidth in more favorable conditions.

\section*{Acknowledgement}
This work was supported by NICT (Grant No. 03801), Japan. Additionally, the authors would like to express their gratitude to PEI Xiaohong of Qtrun Technologies for providing Network Signal Guru (NSG) and AirScreen, the cellular network drive test software used for result collection and analysis in this research. Finally, we gratefully thank every member of Melodious Chorus for the inspiration and for driving our research using the power of \emph{utaite}, among them ElimZ, Qualia, Moonlighteas, Starlightloverful, Ryuuen, kuai, PatzCoject, TAMichi, AbenDroTz, Hazel, Utako, AnatsuKun, EnGz, Dusk, crin, Bookiezz, Amaru, RoBiNET, teajar, Muai, Ushikun, Mai-chan, Miha-tan, FaiiiRy, REwiew, ViewVY, CChika, kairyuramon, Bloodyflora, YellowHaya, U\_na, Toku, Nekonene, kanaseki, LizzyLLJUNG, Saki, Itz\_FourClover, JukkaWink, SokazeNiji, Wasenshi, AmeTsun, and Shaun.




%
\Urlmuskip=0mu plus 1mu\relax
\bibliographystyle{IEEEtran}
\bibliography{IEEEabrv,reference}

\end{document}